# On the Estimate Measurement Uncertainty of the Insertion Loss in a Reverberation Chamber Including Frequency Stirring

Angelo Gifuni, *Member, IEEE*, Luca Bastianelli, Maurizio Migliaccio, *Fellow, IEEE,* Franco Moglie, *Senior Member, IEEE*, Valter Mariani Primiani, *Senior Member, IEEE*, and Gabriele Gradoni, *Member, IEEE*

*Abstract*— In this paper, it is shown an enhancement of a previous model on the measurement standard uncertainty (MU) of the insertion loss (IL) in a reverberation chamber (RC) including frequency stirring (FS). Differently from the previous model, the enhanced does not require specific conditions on the parameter to be measured. Such an enhancement is applicable for all usable measurement conditions in RCs. Moreover, a useful majorant is also shown; it is obtained under a weak condition on the coefficient of variation (CV) of the parameter to be measured. Results by measurements support the validity of the proposed enhancement and of the majorant.

## I. Introduction

Measurement uncertainty (MU) quantification is very important to improve the applications of reverberation chambers (RCs) [1]. Hybrid stirring increases the number of uncorrelated samples and, consequently, reduces the MU [1]-[7]. In this paper, we consider a hybrid stirring as realized by a combination of frequency stirring (FS) and mechanical stirring (MS) [2]-[3]. The FS measurements also allow us the transformation in time domain [8]-[10]. The MU of the insertion loss (IL) in an RC with hybrid MS and FS was addressed in [11], where a model was developed and achieved under conditions of well-stirred fields; it is here called previous model. In [11], MU is estimated following the approach described in [12], considering it as a type A uncertainty. This type of uncertainty is normally the main component for MU in RCs [13]. The type B evaluation uncertainty depends on Manufacturer's specifications of the instrumentations, as well as on the specific calibration procedure used for measurements, which can change from case to case; however, the instrumentation used for measurements and main concerning settings will be also shown here. Actually, a wider treatise on the MU in RCs was addressed in [13], where an approach similar to that in [11] was used, as it will be specified below; nevertheless, some meaningful differences in the approaches should be discussed. We will discuss that below in the section V. The purpose of this paper is to enhance the previous model and the concerning usability. The enhanced model does not require specific conditions for its validity; it is de facto a generalization of the previous model. It is found that such an enhancement is applicable for all usable measurement conditions of IL in RCs including conditions at low frequencies. A useful majorant of the standard MU is also obtained; it requires that the coefficient of variation (CV) of the measured samples be less than or equal to one. We find that the previous model is the same as the majorant. It can be applied when a conservative margin for statistical fluctuations is considered and the abovementioned CV is less than one.

## II Theory

We develop the enhancement by considering the IL as made in [11]. We can write [11]:

$$IL = \left\langle \left|S_{21}\right|^2 \right\rangle_N = \left\langle E^2 \right\rangle_N, \quad (1)$$

where $<\ >_N$ represents the ensemble average with respect to the $N$ uncorrelated field configurations from MS of metallic stirrer(s) in the chamber. $E^2$ represents the squared amplitude of the transmission coefficient $S_{21}$; it is a random variable (RV). Actually, $IL$ is a sample mean (SM) and therefore has statistical fluctuations: it is an RV. We can write the mean, variance, and CV of the RV $IL_f$, respectively, as follows[1] [11]:

$$\mu_{IL_f} = IL_{f,0}, \quad (2)$$

$$\sigma^2_{IL_f} = \frac{\sigma^2_{E^2,f,0}}{N}, \quad (3)$$

$$\delta^2_{IL_f} = \frac{\sigma^2_{IL_f}}{IL^2_{f,0}}, \quad (4)$$

Manuscript received May 12, 2018.
A. Gifuni, M. Migliaccio, and S. Perna are with the Dipartimento di Ingegneria, Università di Napoli Parthenope, Centro Direzionale di Napoli, Napoli 80143, Italy (e-mail: angelo.gifuni@uniparthenope.it; maurizio.migliaccio@uniparthenope.it).
L. Bastianelli, F. Moglie, and V.M. Primiani are with the Dipartimento di Ingegneria dell'Informazione, Università Politecnica delle Marche, 60131 Ancona, Italy (e-mail: f.moglie@univpm.it; l.bastianelli@pm.univpm.it; v.mariani@univpm.it).
G. Gradoni is with the School of Mathematical Sciences and with the George Green Institute for Electromagnetics Research, Department of Electrical and Electronics Engineering, University of Nottingham, Nottingham NG7 2RD, U.K. (e-mail: gabriele.gradoni@nottingham. ac.uk).

---

[1] Differently from [11], here, the mean squared of the IL, as well as other squared means, is written with no brackets.

where $N$ is the number of uncorrelated samples used to estimate the SM $IL_f$, $f$ is the frequency, and $\sigma^2_{E^2,f,0}$ is the variance of $E^2$. In order to analyze the behavior of the enhanced model, we will use the following conditions:

$$\delta_{E^2,f,0} = \frac{\sigma_{E^2,f,0}}{\mu_{IL_f}} = \frac{\sigma_{E^2,f,0}}{IL_{f,0}} = \text{constant in FSB}, \quad (5a)$$

$$\delta_{E^2,f,0} = \frac{\sigma_{E^2,f,0}}{\mu_{IL_f}} = \frac{\sigma_{E^2,f,0}}{IL_{f,0}} \leq 1. \quad (5b)$$

$$\delta_{E^2,f,0} = \frac{\sigma_{E^2,f,0}}{\mu_{IL_f}} = \frac{\sigma_{E^2,f,0}}{IL_{f,0}} > 1. \quad (5c)$$

Note that for well-stirred fields, we have $\mu_{IL_f} = IL_{f,0} = \sigma_{E^2,f,0}$; but, in general, it turns out that $\mu_{IL_f} = IL_{f,0} \geq \sigma_{E^2,f,0}$, i. e., the condition (5b) is met in most cases for fields in RCs; often, both the conditions (5a) and (5b) are met. However, at the low frequencies, both the exponential and the Rician distribution could be unmet, as well as both conditions (5a) and (5b) could be unmet. We want to develop a model valid under any condition (5a)-(5c). When the condition (5b) is met the IL includes a direct component; such a direct component can be both desirable [14] and undesirable; the latter is the typical case of RCs loaded [15]-[16]. When the samples are acquired both by mechanical and frequency stirring[2], then $IL_f$ is denoted by $IL_{N,\Delta f}$; we can write [11]:

$$IL_{N,\Delta f} = W = \left\langle \left\langle E^2 \right\rangle_N \right\rangle_k = IL_{\Delta f, N} = \left\langle \left\langle E^2 \right\rangle_k \right\rangle_N, \quad (6)$$

where the subscript $\Delta f$ means that the averages are made over $k$ uncorrelated frequency samples in FSB, which is denoted by $\Delta f$ [11]. Here, we consider the averages with respect to $N$ first and then those with respect to $k$ [11]; it is implicit that we consider step-tuned RCs [17]-[18]. The averages for each frequency point correspond to SMs including only the MS from to metallic stirrer(s). Such SMs are assumed to be uncorrelated RVs and they are denoted by $IL_{f1}, IL_{f2}, \cdots, IL_{fk}$. Their corresponding mean values are denoted by $IL_{f1,0}, IL_{f2,0}, \cdots, IL_{fk,0}$. The RV $W$ given by (6) can be expressed as follows [11]:

$$W = \frac{1}{k}\left[ IL_{f1} + IL_{f2} + \cdots + IL_{fk} \right]. \quad (7)$$

Note that $\Delta f = f_k - f_1$, where $f_1$ and $f_k$ are the minimum and the maximum frequency of the FS. We are interested in the mean and variance of $W$. We can write:

$$\mu_W = W_0 = \frac{1}{k}\left[ IL_{f1,0} + IL_{f2,0} + \cdots + IL_{fk,0} \right], \quad (8)$$

$$\sigma_W^2 = \frac{1}{kN}\left\{ \frac{\sigma^2_{E^2,f1,0} + \sigma^2_{E^2,f2,0} + \cdots + \sigma^2_{E^2,fk,0}}{k} \right\}. \quad (9a)$$

We want to transform (9a) so that it gives a significant connection between MS and FS. We can write:

---

[2] The MS considered in (6) is limited to metallic stirrer(s) again.

$$\sigma_W^2 = \frac{1}{kN}\left\{ \begin{array}{l} \dfrac{\delta^2_{E^2,f1,0}IL^2_{f1,0} + \delta^2_{E^2,f2,0}IL^2_{f2,0}}{k} \\ + \cdots + \delta^2_{E^2,fk,0}IL^2_{f2,0} \\ \hline k \end{array} \right\}, \quad (9b)$$

where $\delta^2_{E^2,f,0} = \sigma^2_{E^2,f,0}/IL^2_{f,0}$. The mean $IL^2_{fi,0}$ ($i = 1, 2, \ldots, k$) changes as the frequency changes; the variations depend on the RC and FSB. The means $\delta^2_{E^2,f,0}$ and $IL^2_{f,0}$, as well as the corresponding sample estimates, can be thought how two RVs, whose values are associated by the frequencies $f_i$. We can write [19], [20]:

$$\left\langle \delta^2_{E^2,f,0} IL^2_{f,0} \right\rangle_k = \delta^2_{E^2,\Delta f,0} IL^2_{\Delta f,0} + \left[ Cov\left(\delta^2_{E^2,f,0}, IL^2_{f,0}\right) \right]_k, \quad (10a)$$

where $\delta^2_{E^2,\Delta f,0}$ and $IL^2_{\Delta f,0}$ are the means of $\delta^2_{E^2,f,0}$ and $IL^2_{f,0}$ in the FSB, respectively; $Cov$ means covariance; the subscript $k$ means that the concerning parameter is referred to the FSB. The covariance is equal to zero when the RVs are uncorrelated or when (5a) is met; in the latter case, we can write: $\delta^2_{E^2,0} = \text{const.} = \delta^2_{E^2,\Delta f,0}$. The RVs $\delta^2_{E^2,f,0}$ and $IL^2_{f,0}$, which are estimated by corresponding sample means from $N$ uncorrelated sampling data of $S_{21}$, are never totally uncorrelated as the former includes an effect of the latter. However, they are sufficiently uncorrelated, so that (10a) can be well approximated as follows:

$$\left\langle \delta^2_{E^2,f,0} IL^2_{f,0} \right\rangle_{FSB} = \delta^2_{E^2,\Delta f,0} IL^2_{\Delta f,0}. \quad (10b)$$

It is highlighted that (10b) is valid also in case of sample estimates. We can also write [11]:

$$W_0^2 + \sigma^2_{\Delta f,0} = \left[ \frac{IL^2_{f2,0} + IL^2_{f2,0} + \cdots + IL^2_{f2,0}}{k} \right], \quad (11)$$

where $\sigma_{\Delta f,0}$ is the standard deviation of the means $IL_{f1,0}, IL_{f2,0}, \cdots, IL_{fk,0}$. Manipulating (9), (10b), and (11), we can write:

$$\sigma_W = \frac{W_0}{\sqrt{kN}}\left(\delta^2_{E^2,\Delta f,0}\right)^{1/2}\sqrt{1+\delta^2_{\Delta f,0}}, \quad (12)$$

$$\delta_W = \left(\delta^2_{E^2,\Delta f,0}\right)^{1/2}\frac{\sqrt{1+\delta^2_{\Delta f,0}}}{\sqrt{kN}}, \quad (13)$$

where $\sigma_W$ and $\delta_W$ are the standard MU and the relative standard MU of $W$, respectively; the CV $\delta_{\Delta f,0} = \sigma_{\Delta f,0}/W_0$. When (5a) is met and $\delta_{E^2,0} = 1$, which corresponds to the case of well-stirred fields, (12) and (13) become equal to (10) and (13) in [11], respectively, as expected. Practically, $\sigma_W$ and $\delta_W$ are also RVs as parameters on the right side of (12) and (13), as well as those in similar eqs. below, are sample estimates. They depend however on $N$; in these cases, we omit the zero at their subscript. When (5a) is met, a variation of the enhanced model (12)-(13) can be obtained; in fact, (12) and (13) became as follows [21]:

$$\sigma_W = \frac{W_0}{\sqrt{kN}}\delta_{E^2,\Delta f,0}\sqrt{1+\delta^2_{\Delta f,0}}, \quad (14)$$

$$\delta_W = \delta_{E^2,\Delta f,0} \frac{\sqrt{1+\delta_{\Delta f,0}^2}}{\sqrt{kN}}, \quad (15)$$

where $\delta_{E^2,\Delta f,0} = \delta_{E^2,f,0}$ for the assumption (5a). Since population parameters are estimated by the corresponding sample statistics[3], which uses $N$ uncorrelated sampling data of $S_{21}$, we can de facto know if (5a) is met only when $N$ is much greater than one; in fact, in such cases, the statistical fluctuations are very reduced. When $N$ is not much greater than one, we can assume that (5a) is met and calculate its average in the FSB; the comparison of the results with those from measurements proves if the assumption was true. Note that $\delta_{E^2,\Delta f,0}$ and $\left(\delta_{E^2,\Delta f,0}^2\right)^{1/2}$ are mean and root mean square (RMS) of $\delta_{E^2,f,0}$ in the FSB. They tend to the same value when the variance of $\delta_{E^2,f,0}$ or of the concerning sample estimate tends to zero in the FSB. In section IV, it is shown that when $N$ is greater than or equal to eight, (12)-(13) practically give the same results of (14)-(15). Differently, results from (14)-(15) are worst; in particular, the concerning standard MU and relative standard MU are smaller than the corresponding from (12)-(13) as expected. It will be shown that results from (12)-(13) match those from measurements. Moreover, on equal $N$ value, the difference between results from (12)-(13) and those from (14)-(15) is maximum when the $K$-Factor is zero. By measurements, which are the samples of $S_{21}$ taken in the RC at frequencies $f_i$ ($i = 1, 2, …, k$) within the FSB, one can estimate the means $IL_{f1,0}, IL_{f2,0}, \cdots, IL_{fk,0}$, as well as $W_0^2$, $\delta_{E^2,\Delta f,0}$, and $\delta_{\Delta f,0}^2$. The variances are estimated as sample variances. Then, by using (12)-(15), we can calculate the corresponding standard and relative standard MUs. For $k = 1$, the achieved models retrieve the pure MS model of (3) and (4), of which they are extensions. Since the means $IL_{f1,0}, IL_{f2,0}, \cdots, IL_{fk,0}$ are estimated by the corresponding sample means, their statistical fluctuations increase with the decrease of $N$; in particular, both $\delta_{E^2,\Delta f,0}$ and $\left(\delta_{E^2,\Delta f,0}^2\right)^{1/2}$ are appreciably underestimated when $N$ is small [13]. It will be confirmed by results in section IV. By following assumptions and developments made in [11, after eq. 14], we can write:

$$\sigma_{W_{mp}}^2 = \delta_{E^2,\Delta f,0}^2 \frac{W_{mp,0}^2}{pkN}\left(1+\delta_{sp,p,0}^2\right)\left(1+\delta_{\Delta f,0}^2\right) + \frac{\sigma_{sp,p,0}^2}{p}, \quad (16)$$

where the subscripts $p$, $mp$, and $sp$ mean $p$ independent positions of at least one of the two antennas, multiple positions, and a single position, respectively; $\sigma_{sp,p,0}^2$ is the variance due to the lack of perfect uniformity [11]. Note that the constancy of $\delta_{E^2,\Delta f,0}$ for all positions $p$ is an assumption

---

[3] This is the reason for which the symbol $\delta_{E^2,\Delta f,0}$ is used in (14) and (15) instead of $\delta_{E^2,f,0}$.

absolutely acceptable. If $k = 1$ (only MS), then (16), becomes as follows:

$$\sigma_{W_{mp}}^2 = \delta_{E^2,f,0}^2 \frac{W_{mp,0}^2}{pN}\left(1+\delta_{sp,p,0}^2\right) + \frac{\sigma_{sp,p,0}^2}{p}. \quad (17)$$

It is useful to write (16) as follows:

$$\sigma_{W_{mp}}^2 = \delta_{E^2,\Delta f,0}^2 \frac{W_{mp,0}^2}{pkN} CF + \frac{\sigma_{sp,p,0}^2}{p}, \quad (18)$$

where

$$CF_0 = \left(1+\delta_{\Delta f,0}^2\right)\left(1+\delta_{sp,p,0}^2\right). \quad (19)$$

It is also useful to write the ratio $R_{\delta sq,0} = \delta_{sp,p,0}^2/\delta_{\Delta f,0}^2$. If $R_{\delta sq,0} \ll 1$, then $CF_0 \cong \left(1+\delta_{\Delta f,0}^2\right)$. We can write:

$$\sigma_{W_{mp}} = \sqrt{\delta_{E^2,\Delta f,0}^2 \frac{W_{mp,0}^2}{pkN} CF_0 + \frac{\sigma_{sp,p,0}^2}{p}} = \sqrt{\sigma_1^2 + \sigma_2^2}, \quad (20)$$

where

$$\sigma_1 = \delta_{E^2,\Delta f,0} \frac{W_{mp,0}}{\sqrt{pkN}} \sqrt{CF_0}, \quad (21)$$

$$\sigma_2 = \sigma_{sp,p,0}/\sqrt{p}. \quad (22)$$

Equations (21) and (22) allow to estimate standard MU contributions $\sigma_1$ and $\sigma_2$; however, they are not completely uncorrelated [11]. It is important to note that the value of $\delta_{E^2,\Delta f,0}$ particularly affects $\sigma_1$; that is, it particularly affects (12)-(15). The total relative MU can be written as follows:

$$\delta_{W_{mp}} = \sqrt{\delta_{E^2,\Delta f,0}^2 \frac{CF_0}{pkN} + \frac{\sigma_{sp,p,0}^2}{W_{mp,0}^2 p}}$$
$$= \sqrt{\delta_{E^2,\Delta f,0}^2 \frac{CF_0}{pkN} + \frac{\delta_{sp,p,0}^2}{p}} = \sqrt{\sigma_{1,r}^2 + \sigma_{2,r}^2}, \quad (23)$$

where $\sigma_{1,r}^2$ and $\sigma_{2,r}^2$ are the contributions to the relative MU, which correspond to the uncertainties squared $\sigma_1^2$ and $\sigma_2^2$, respectively.

Note that if (5b) is met, we can write:

$$\sigma_{W_c} \leq \frac{W_{c,0}}{\sqrt{kN}} \sqrt{1+\delta_{c,\Delta f,0}^2}, \quad (24)$$

$$\delta_{W_c} \leq \frac{\sqrt{1+\delta_{c,\Delta f,0}^2}}{\sqrt{kN}}. \quad (25)$$

Note that the right sides of (24) and (25) are the same as in the previous model, and they give majorants of the corresponding standard MUs. It is specified that the subscript $c$ in (24) and (25) denote that fields meet the condition (5b). It is important to highlight that (12) and (13), as well as the corresponding (20) and (23), are a general model for the standard MU of the IL in RCs; in fact, (10)-(13) in [11], as well as (24) and (27) in [11], are a particular case of (12)-(13) and (20) and (23), respectively, which occurs when $\delta_{E^2,0} = 1$. Finally, before we show the measurement setup, it is useful to express the CV $\delta_{E^2,f,0}$ by the *K*-Factor, which is denoted by $K_{f,0}$; we can write:

$$\delta_{E^2,f,0} = \sqrt{\frac{2K_{f,0}+1}{2K_{f,0}+1+K_{f,0}^2}}, \qquad (26)$$

where

$$K_{f,0} = \left(\mu_{1,f,0}^2 + \mu_{2,f,0}^2\right)/2\sigma_{E^2,f,0}^2, \qquad (27)$$

$\mu_{1,f,0}$ and $\mu_{2,f,0}$ are the means of the real and imaginary part, respectively, of the coefficient $S_{21}$, and $\sigma_{E^2,f,0}$ is the common standard deviation of such parts. When $S_{21}$ has a Rician distribution, (26) is less than one. From (26), one notes that if $K_{f,0}$ is constant in the FSB then also $\delta_{E^2,f,0}$ is constant.

### III. MEASUREMENTS SETUP

Measurements are made in the RC at Università Politecnica delle Marche, Ancona, Italy, which works in step mode for measurements used in this paper. The measurement setup and acquisition settings are the same as in [11], except that in this case two type of configurations of the antennas are used for measurements: one configuration minimizes the direct link between the antennas, which are distant and cross-polarized, and the other one maximizes it. In the latter case, the antennas are on the line of sight at a known distance each other; they are tip-to-tip positioned and co-polarized; several distances are used for measurements but only results concerning the distances of 0.05 m and 0.3 m are shown for shortness. The former and latter measurement configurations are here called A and B, respectively. It is specified that the measurement setup includes a four-port VNA, model Agilent 5071B, and two antennas, model Schwarzbeck Mess-Elektronik USLP 9143, whose usable frequency range (FR) ranges from 250 MHz to 7 GHz for EMC tests. The IF bandwidth and source power, which determine the instrument measurement uncertainty along with the set FR and amplitude of the measured transmission coefficient, are set to 3 kHz and 0 dBm, respectively. Over the FR from 0.2 GHz to 8.2 GHz, 16,001 frequency points are acquired with a step frequency (SF) of 500 kHz for a number of mechanical positions $M = 64$ [11]. Note that the number 64 corresponds to the total number of acquired stirrer positions, which in turns corresponds to the total number of acquired (frequency) sweeps ($M = 64$) [11]. The total sweeps are divided in *n* sets of (frequency) sweeps, so that each set includes *N* sweeps and $M = n \cdot N$. The settings *n* and *N* can be changed to test the enhanced model [11]. For each sweep, the total number of processed frequency points $\kappa = 16,000$ is divided in *q* sets of frequencies, so that $\Delta f = (k - 1) \cdot SF$ and $\kappa = k \cdot q$. Unlike what was made in [11], the symbol for the total frequency points is here denoted by $\kappa$ to avoid confusion with the symbol of the *K*-Factor. The value of *q* is the number of FSB or $\Delta f$ included in the FR. The mean $W_0$ in (12) is estimated *n* times and the standard deviation of such *n* averages $W_i$ ($i = 1, 2, \cdots, n$) is calculated [11]. The calculated standard deviation is an estimate of the measured standard uncertainty. When such an uncertainty is normalized to the average of the averages $W_i$, an estimate of the relative standard uncertainty is obtained. The measured standard MU is compared to the corresponding expected standard MUs, which are obtained by applying (12), as well as (14), and (24). They are applied by using any of the *n* estimates $W_i$ and the corresponding estimates of $\sigma_{\Delta f,0}^2$, $\left(\delta_{E^2,\Delta f,0}^2\right)^{1/2}$, and $\delta_{E^2,\Delta f,0}$; clearly, the estimates of $\left(\delta_{E^2,\Delta f,0}^2\right)^{1/2}$ and $\delta_{E^2,\Delta f,0}$ are also calculated *n* times, and they are used in (12) and (14), respectively, as mentioned above. Similarly, the measured relative standard MU is compared to the corresponding expected relative standard MUs, which are obtained by applying (13), (15), and (25). The non-correlation of samples is verified by autocorrelation function (ACF). Here, the threshold used is 1/*e*, where *e* is the Neper's number. In general, thresholds of 0.5 and 0.7 could be also used [22]; however, the higher the threshold the higher the residual correlation of samples [21]. Note that the 64 frequency sweeps of each IL measurement can be thought as a matrix of 64 rows and 16,001 columns, where along each row changes only the frequency (FS) whereas along each column change only the stirrer position (MS). The ACF is calculated for both any row and column. For measurements where the IL includes a significant variable direct component, the ACF is considerably affected. A short sequence of frequency samples, where the average of the direct component is removed, could be considered, as made in [21]; this method has the drawback to use only a few samples for the estimate of the ACF and, however, it is not reliable [21]; therefore, it is not used here. Here, the direct component is removed before calculating the ACF for measurements concerning the configuration B; it is removed for each frequency point, i.e., it is removed for both MS and FS. The direct component to be removed is obtained by using all 64 sweeps. For both measurements from configuration A, where it is not necessary to remove the residual direct component, and measurements from configuration B, acceptable results are obtained according to the abovementioned threshold, which are not explicitly shown here for the sake of shortness. However, to ensure non-correlated samples in all the FR and for any FSB, a decimation of samples from 1 through 8 is made for samples concerning the configuration B. Similarly, a decimation of samples from 1 through 2 is made for samples concerning the configuration A. Hence, SF becomes 1 MHz for configuration A measurements and 4 MHz for configuration B measurements. Finally, we note that when an appreciable direct component is not present or when it is removed and the stirred component is well stirred, the non-correlation can be verified by using the correlation coefficient (CC) applied to the amplitude squared of samples [23]. By using such a method, it is confirmed that results worse when the FSB increases, as well as when it is too much small, according to the number of samples [23].

### IV. RESULTS

The effect of the enhancement of the previous model and of the majorant is well-visible in $\sigma_1$. Therefore, in order to make effective and simple the verification of the proposed models, we use (12)-(13), (14)-(15), and (24)-(25). Fig. 1 and 2 show the standard MUs and the relative standard MUs given by (12), (14), (24), and (13), (15), (25), respectively, for measurements concerning the configuration A. Note that $\Delta f = (k - 1) \cdot 1$ MHz. Fig. 3 shows an enlargement of Fig. 2 at low frequencies. From Fig. 4 to Fig. 7, the CV $\delta_{\Delta f}$, the CV $\delta_{E^2,f}$ along with its average value and RMS value, and $K$-Factor are shown; in particular, Fig. 6 shows an enlargement of Fig. 5 at low frequencies. All processing settings ($N$, $k$, etc.) are shown in the concerning captions of Figs.

in the FSB; for all traces, $M = 64$, $\kappa = 8,000$, and $N = 8$. $k = 1$ for blue and unmarked trace and $k = 10$ ($\Delta f = 9$ MHz) for the green and red traces, which are cross-marked and circled-marked, respectively.

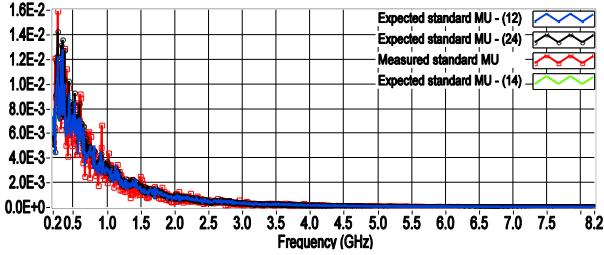

Fig. 1. Standard MU from the configuration A; for measured and expected standard uncertainties, $M = 64$, $\kappa = 8,000$, $N = 8$, $n = 8$, and $k = 10$ ($\Delta f = 9$ MHz).

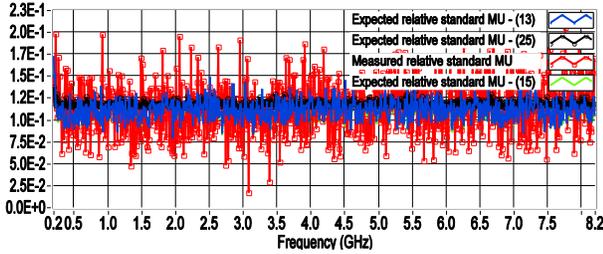

Fig. 2. Relative standard MU from the configuration A; for measured and expected relative standard uncertainties, $M = 64$, $\kappa = 8,000$, $N = 8$, $n = 8$, and $k = 10$ ($\Delta f = 9$ MHz).

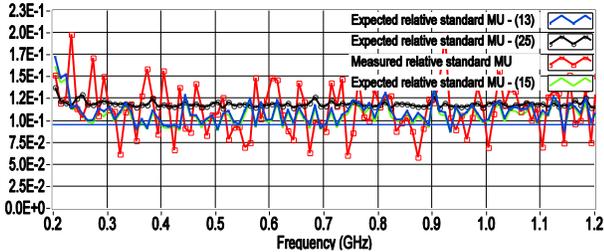

Fig. 3. Enlargement of the Fig. 2 at low frequencies.

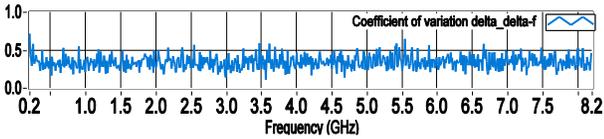

Fig. 4. CV $\delta_{\Delta f}$ from the configuration A; $N = 8$, $k = 10$ (9 MHz).

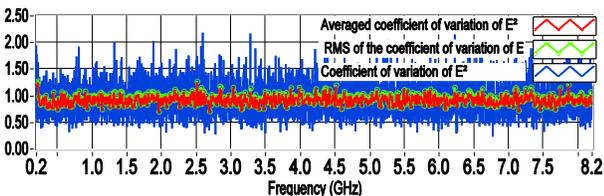

Fig. 5. CV $\delta_{E^2,f}$ from the configuration A, its average value, and RMS value

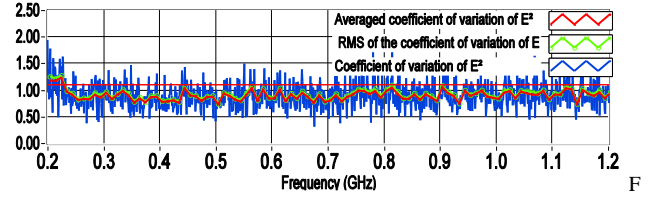

Fig. 6. Enlargement of the Fig. 5 at low frequencies.

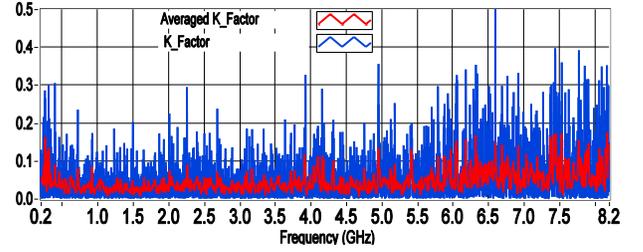

Fig. 7. $K$-Factor from the configuration A; for both traces $M = 64$, $\kappa = 8,000$, and $N = 64$. $k = 1$ for blue and unmarked trace and $k = 10$ ($\Delta f = 9$ MHz) for red and cross-marked trace.

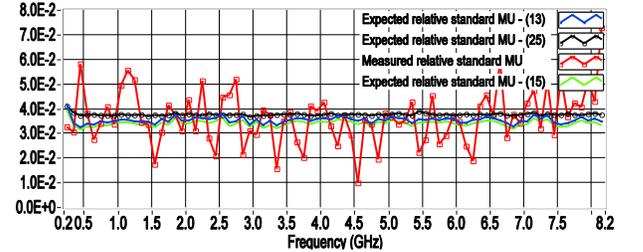

Fig. 8. Relative standard MU from the configuration A; for measured and expected relative standard uncertainties, $M = 64$, $\kappa = 8,000$, $N = 8$, $n = 8$, and $k = 100$ ($\Delta f = 99$ MHz).

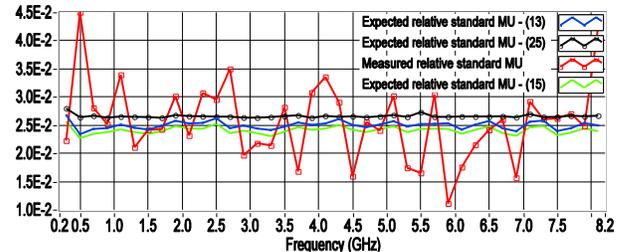

Fig. 9. Relative standard MU from the configuration A; for measured and expected relative standard uncertainties, $M = 64$, $\kappa = 8,000$, $N = 8$, $n = 8$, and $k = 200$ ($\Delta f = 199$ MHz).

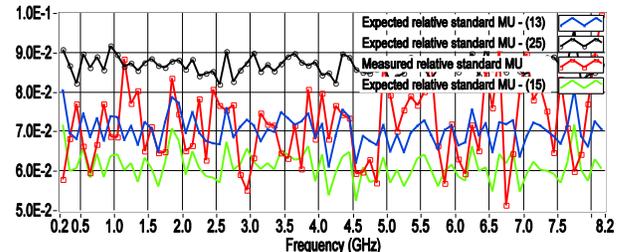

Fig. 10. Relative standard MU from the configuration A; for measured and expected relative standard uncertainties, $M = 64$, $\kappa = 8,000$, $N = 2$, $n = 32$, and $k = 100$ ($\Delta f = 99$ MHz).

All Figs show the concerning statistical fluctuations. The comparison between measured standard MUs and corresponding expected standard MUs shows that (12)-(13), as well as (14)-(15) are supported by measurement results. In order to prove that the models works well also for different FSBs, expected relative standard MSs are shown in Fig. 8 and Fig. 9, where $k = 100$ ($\Delta f = 99$ MHz) and $k = 200$ ($\Delta f = 199$ MHz), respectively. It is also confirmed that expected results from (24) and (25) are the same as those from (12) and (13), respectively, when $K = 0$, which implies $\delta_{E^2} = 1$ (the equal sign has to be taken in (24)-(25)), except at the low frequencies ($f < 250$ MHz), where a deviation is expected and observed (see Fig. 3). By Figs. (1)-(9), it is also noted that (12)-(13) and (14)-(15) give practically the same results for $N = 8$. In Fig. 10, where $N$ is 2, it is well visible the difference between results from (13) and (15). Such a difference is due to the statistic fluctuations, which increase as $N$ decreases, as mentioned above; the same applies to (12) and (14). The slight difference between results from (13) and (25), which is visible in Fig. 10, as well as those between results from (12) and (24), when in (24)-(25) the equal sign has to be taken, is due to the $N$ value; it decreases as $N$ increases because both $\delta_{E^2,\Delta f,0}$ and $\left(\delta^2_{E^2,\Delta f,0}\right)^{1/2}$ are underestimated when $N$ is small, as mentioned above. It is important to note that the measured standard MU and the expected standard MU from (12), as well as the corresponding relative standard MUs, match also when $N$ is small ($N < 4$) for the effect of such an underestimate (see Fig. 10 for the relative standard MUs); otherwise, the abovementioned difference is acceptable from $N = 4$ [11]. Figs. from 11 to 15 show results of measurements concerning the configuration B for d = 0.05 m. In particular, Figs 11 and 12 show expected standard MUs and expected relative MUs along with the corresponding measured MUs. The FSB is 96 MHz. One notes that expected and measured results match again. Note that (24) and (25) are clearly majorants of the corresponding measured uncertainties in these cases.

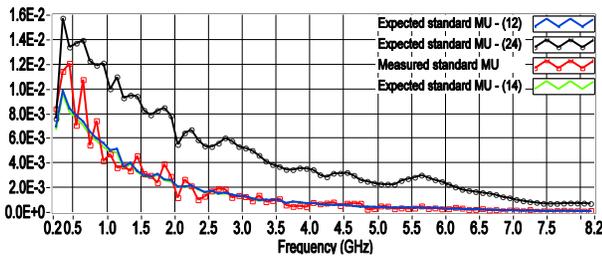

Fig. 11. Standard MU from the configuration B, d = 0.05 m; for measured and expected standard uncertainties, $M = 64$, $\kappa = 2,000$, $N = 8$, $n = 8$, and $k = 25$ ($\Delta f = 96$ MHz).

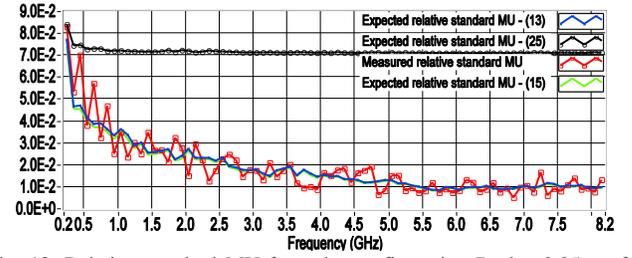

Fig. 12. Relative standard MU from the configuration B, d = 0.05 m; for measured and expected standard uncertainties, $M = 64$, $\kappa = 2,000$, $N = 8$, $n = 8$, and $k = 25$ ($\Delta f = 96$ MHz).

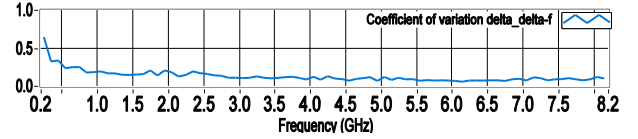

Fig. 13. CV $\delta_{\Delta f}$ from the configuration B, d = 0.05 m; $N = 8$, $k = 25$ (96 MHz).

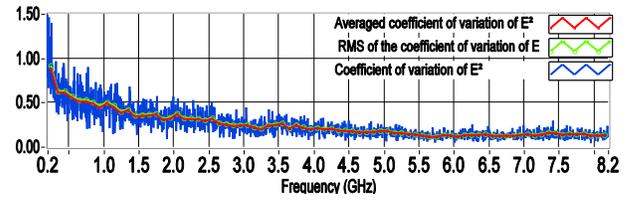

Fig. 14. CV $\delta_{E^2,f}$ from the configuration B (d = 0.05 m), its average value, and RMS value in the FSB; for all traces, $M = 64$, $\kappa = 8,000$, and $N = 8$. $k = 1$ for blue and unmarked trace and $k = 25$ ($\Delta f = 96$ MHz) for the green and red traces, which are cross-marked and circled-marked, respectively.

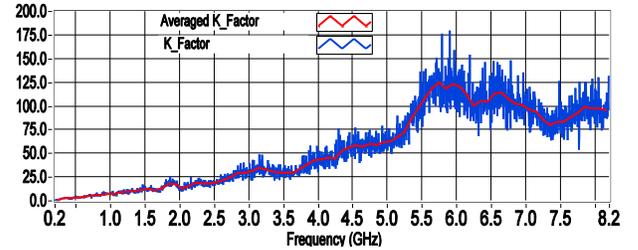

Fig. 15. K-Factor from the configuration B, d = 0.05 m; for both traces $M = 64$, $\kappa = 2,000$, and $N = 64$. $k = 1$ for blue and unmarked trace and $k = 25$ ($\Delta f = 96$ MHz) for red and cross-marked trace.

Figs. (16)-(20) show results of measurements concerning the configuration B for d = 0.3 m.

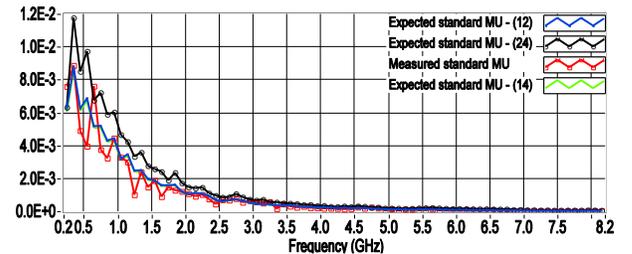

Fig. 16. Standard MU from the configuration B, d = 0.3 m; for measured and expected standard uncertainties, $M = 64$, $\kappa = 2,000$, $N = 8$, $n = 8$, and $k = 25$ ($\Delta f = 96$ MHz).

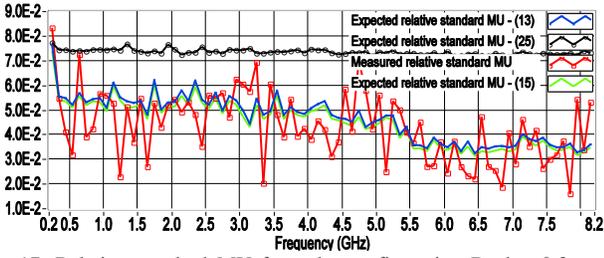

Fig. 17. Relative standard MU from the configuration B, d = 0.3 m; for measured and expected standard uncertainties, $M = 64$, $\kappa = 2{,}000$, $N = 8$, $n = 8$, and $k = 25$ ($\Delta f = 96$ MHz).

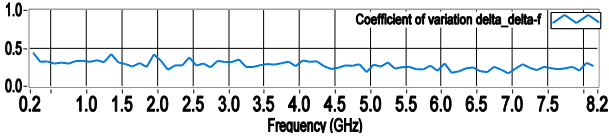

Fig. 18. CV $\delta_{\Delta f}$ from the configuration B, d = 0.3 m; $N = 8$, $k = 25$ (96 MHz).

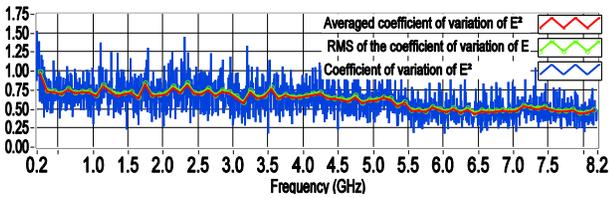

Fig. 19. CV $\delta_{E^2, f}$ from the configuration B (d = 0.03 m), its average value, and RMS value in the FSB; for all traces, $M = 64$, $\kappa = 8{,}000$, and $N = 8$. $k = 1$ for blue and unmarked trace and $k = 25$ ($\Delta f = 96$ MHz) for the green and red traces, which are cross-marked and circled-marked, respectively.

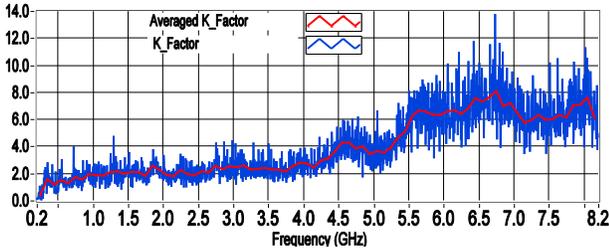

Fig. 20. $K$-Factor from the configuration B, d = 0.3 m; for both traces $M = 64$, $\kappa = 2{,}000$, and $N = 64$. $k = 1$ for blue and unmarked trace and $k = 25$ ($\Delta f = 96$ MHz) for red and cross-marked trace.

It is important to note that results from (12)-(13) are essentially the same as those from (14)-(15) except for $N < 8$ as Fig. 10 shows. However, we consider ultimately the enhanced model (12)-(13) even though we believe that the variation (14)-(15) can generally be used for $N \geq 8$. Finally, we highlight that results in [21, Figs. 22-25], where no decimation was applied, did not match well because samples were partially correlated. The effect of a residual correlation is also appreciable in [21, Fig. 9] for $f > 5$ GHz.

## V. DISCUSSION

The standard MU of the IL of an RC, as well as the relative standard uncertainty, is estimated for type A evaluation uncertainty; they are compared to the corresponding measured uncertainties. The estimate of the MU is made so that the uncertainty component $\sigma_2$ due to the non-uniformity of the field in the RC is highlighted and separately obtained, except the multiplying factor $\left(1+\delta_{sp,p,0}^2\right)$ present in $\sigma_1$. The non-uniformity is affected by the load in the RC and it increases as the load increases. Such a component of uncertainty is connected to the reciprocal location, orientation, and polarization of the transmitting and receiving antennas for a given RC. The model gives good results at low frequencies as well. The non-uniformity of the field in an RC, which is estimated by $\sigma_2$ can not be neutralized by the increase of the samples $N \cdot k$, even though, a marginal reduction of such a component of MU could be achieved by a widening of the FSB [11]. This aspect is very important when $\sigma_2$ has to be reduced. It could be the case where the effect of a strong load on the uniformity has to be reduced or when a very low total uncertainty is necessary. In [13], the PDFs of the interest sample statistics are theoretically achieved; the theory is applied to $\chi^2$ parent distributions with two or six degrees of freedom according to the sample statistic to be processed. The RVs, which are represented by the same amount of samples $N \cdot k \cdot p$ from hybrid stirring, are all assumed to be identically distributed (ID), so that the theoretical PDF is achieved, as well as the concerning uncertainty. It is specified that $M$ in [13] corresponds to $k$ in [11] and here, when MS and FS, but no position stirring, is considered whereas $M$ corresponds to the product $k \cdot p$ in [11] and in this paper when MS, FS, and position stirring are considered. It is important to highlight that the standard MU obtained here and in [11] is equivalent to that obtained in [13] for the average power, when the dependence on the frequency and the non-uniformity of the field are negligible in the FSB. For such measurement conditions, the averages $W$ and $W_{mp}$ certainly exhibit PDFs that can be approximated by a Gauss normal sampling distribution, according to the total number of acquired samples $N \cdot k \cdot p$, and the confidence intervals can also be obtained. However, one can note that the $N \cdot k \cdot p$ RVs are not strictly ID as the IL is subject to the non-uniformity of fields inside an RC. Such variations are affected by the load of the RC as mentioned above. In other words, the RVs have all the same PDF type, but they have not strictly the same mean and standard deviation; that is, they are not strictly statistically equivalent. At low frequencies, the distributions of the field and power deviate from those theoretically known; therefore, the theory applied to $\chi^2$, is an approximation at low frequency. From the experimental point of view, when all samples are mixed up together, the real total uncertainty is obtained. However, the assumption of RVs ID simplifies the theoretical developments and is certainly acceptable for small FSB and little non-uniformity of the field in the RC. The theory can also be extended to cases where fields are partially incoherent, i.e., cases where $K > 0$ [13, pag. 31]. In [13], many PDFs of practical interest and the related uncertainties are achieved, as well as the corresponding confidence intervals, including the PDF and the uncertainty of the maximum value for both field and power. The standard "uncertainty of the uncertainty" is also achieved. Results from some applications expected for the standard [1] are also shown. Finally, we believe that the averages $W$ and $W_{mp}$ exhibit PDFs approximately normal in all common usable measurement conditions in RCs including loaded RCs [20], [24]. Similarly, we believe that the assumption of RVs ID made in [13], along with the extension of the theory to cases where $K > 0$, causes an acceptable approximation in all common usable measurement conditions in RCs including loaded RCs.

## VI. CONCLUSIONS

In this paper, an enhancement of the previous model for the standard MU in an RC is shown; it is de facto a generalization of the previous model. Moreover, a useful majorant of the standard MS is shown as well. By results from measurements, it is shown that enhanced model works well for both high and low frequencies. It includes the previous model as an its particular case and does not require specific conditions for its validity. The majorant requires a weak condition on the CV of the parameter to be measured, i.e., it has to be less than or equal to one. The majorant, which just corresponds to the previous model, could be used when the abovementioned CV is less than one and a conservative margin is considered for the statistical fluctuation; however, it could not work well at low frequencies, where the condition for its validity is not guaranteed. Finally, the comparison between the model shown here and that in [13] was discussed; it is concluded that both approaches are practically sound.

## REFERENCES

[1] IEC 61000-4-21, *Electromagn. Compat. (EMC)*, Part 4-21: Testing and measurement techniques – Reverberation chamber test methods, International Electrotechnical Commission, Geneva, Switzerland, 2011.

[2] U. Carlberg, P. S. Kildal, and J. Carlsson, "Numerical study of position stirring and frequency stirring in a loaded reverberation chamber," *IEEE Trans. Electromagn. Compat.*, vol. 51, pp. 12–17, 2009.

[3] D.A. Hill, *Electromagnetic Fields in Cavities: Deterministic and Statistical Theories*. New York: IEEE Press, 2009.

[4] X. Chen "Measurement uncertainty of antenna efficiency in a Reverberation Chamber," *IEEE Trans. Electromagn. Compat.*, vol. 55, pp. 1331-1334, Dec. 2013.

[5] I.D. Flintoft, G.C. R. Melia, M.P. Robinson, J.F. Dawson, and A.C. Marvin, "Rapid and accurate broadband absorption cross-section measurement of human bodies in a reverberation chamber", *IOP Measurement Science and Technology*, vol. 26, no. 6, art. no. 065701, pp. 1-9, May 2015.

[6] A. Gifuni, G. Ferrara, A. Sorrentino, and M. Migliaccio, "Analysis of the measurement uncertainty of the absorption cross section in a reverberation chamber", *IEEE Trans. Electromagn. Compat.*, vol. 57, no. 5, pp. 1262-1265, Oct. 2015.

[7] A. Gifuni, I.D. Flintoft, Simon J. Bale, G.C. R. Melia, and A.C. Marvin, "A theory of alternative methods for measurements of absorption cross section and antenna radiation efficiency using nested and contiguous reverberation chambers", *IEEE Trans. Electromagn. Compat.*, vol. 58, pp. 678-685, June 2016.

[8] C.L. Holloway, H.A. Haider, R.J. Pirkl, W.F. Yong, D.A. Hill, J. Ladbury, "Reverberation chamber techniques for determining the radiation and total efficiency of antennas," *IEEE Trans. Electromagn. Compat.*, vol.60, pp.1758-1770, April 2012.

[9] A. Gifuni and S. Perna, "Analysis on the calculation of the inverse discrete Fourier transform (IDFT) of passband frequency response measurements in terms of lowpass equivalent response," *Prog. In Electromagn. Research*, vol. 160, 63–69, 2017.

[10] X. Zhang, M. Robinson, I.D. Flintoft, and J.F. Dawson "Inverse Fourier transform technique of measuring averaged absorption cross section in the reverberation chamber and Monte Carlo study of its uncertainty," *Electrom. Compat., IEEE Intern. Symp.* EMC Europe, pp. 263-267, 2016, DOI:10.1109/EMCEurope.2016.7739161.

[11] A. Gifuni, L. Bastianelli, F. Moglie, V. M. Primiani, and G. Gradoni, "Base-case model for measurement uncertainty in a reverberation chamber including frequency stirring," *IEEE Trans. Electromagn. Compat.*, DOI: 10.1109/TEMC.2017.2763627.

[12] B. N. Taylor and C. E. Kuyatt, "Guidelines for Evaluating and Expressing the Uncertainty of NIST Measurement Results," NIST Tech. Note 1297, Sept. 1994.

[13] L. R. Arnaut, Measurement Uncertainty in Reverberation Chambers – I. Sample Statistics, 2nd. ed., NPL Report TQE2, pp. 1–136, Dec. 2008.

[14] C. Lemoine, E. Amador, and P. Besnier, "On the *K*-factor estimation for Rician channel simulated in reverberation chamber," *IEEE Trans. Antennas Propag.*, vol. 59, no. 3, pp. 1003–1012, Mar. 2011.

[15] C.L. Holloway, D.A. Hill, J.M. Ladbury, and G. Kapke, "Requirements for an effective reverberation chamber: unloaded or loaded," *IEEE Trans. Electromagn. Compat.*, vol.48, pp.187-194, Feb. 2006.

[16] K.A. Remley, R.J. Pirkl, C.M. Wang, D. Senic, A.C. Homer, M.V. North, M.G. Becker, R.D. Horansky, and C.L. Holloway, "Estimating and correcting the device-under-test transfer function in loaded reverberation chambers for over-the-air-tests," *IEEE Trans. Electromagn. Compat.*, vol. 59, pp. 1724-1734, Dec. 2017.

[17] V. Rajamani, C.F. Bunting, and J.C. West "Stirred-Mode operation of reverberation chambers for EMC testing," *IEEE Trans. Electromagn. Compat.*, vol.61, pp.2759-2764, Oct. 2012.

[18] L.R. Arnaut, "Effect of local stir and spatial averaging on measurement and testing in mode-tuned and mode-stirred reverberation chambers," *IEEE Trans. Electromagn. Compat.*, vol. 43, pp. 305-325, August 2001.

[19] D. Blumenfeld, *Operations Research Calculations Handbook*, CRC Press, New York, 2001.

[20] A. Papoulis, "Probability, random variables and Stochastic Process," New York: McGraw-Hill, 1991.

[21] A. Gifuni, L. Bastianelli, G. Gradoni, M. Migliaccio, F. Moglie, and V. Mariani Primiani, "Applicability of measurement uncertainty models in a reverberation chamber including frequency stirring," in *Proc. IEEE Int. Symp. Electromagn. Compat. Sig. Pow. Integr.* (EMC-SIPI), Long Beach, CA, USA, 2018.

[22] K.A. Remley, J. Dortmans, C. Weldon, R.D. Horansky, T.B. Meurs, C.-M. Wang, D.F. Williams, and C.L. Holloway, "Configuring and verifying reverberation chambers for testing cellular wireless devices," *IEEE Trans. Electromagn. Compat.*, vol. 58, no. 3, pp. 661–671, June, 2016.

[23] L.R. Arnaut, M.I. Andries, J. Soil, P. Besnier "Evaluation method for the probability distribution of the quality factor of mode-stirred reverberation chambers," *IEEE Trans. Electromagn. Compat.*, vol. 43, pp. 305-325, August 2001.

[24] A. Gifuni, A. Sorrentino, G. Ferrara, M. Migliaccio, "An estimate of the probability density function of the sum of a random number *N* of independent random variables," *Journ. of Comput. Engineering, Hindawi Publishing Corporation*, vol. 2015, 12 pages, 2015. doi:10.1155/2015/801652.